\documentclass[conference]{IEEEtran}
\IEEEoverridecommandlockouts

\usepackage{cite}
\usepackage{amsmath,amssymb,amsfonts}
\usepackage{algorithmic}
\usepackage{graphicx}
\usepackage{textcomp}
\usepackage{xcolor}
\def\BibTeX{{\rm B\kern-.05em{\sc i\kern-.025em b}\kern-.08em
    T\kern-.1667em\lower.7ex\hbox{E}\kern-.125emX}}
\begin{document}

\title{An Intelligent and Privacy-Preserving Digital Twin Model for Aging-in-Place\\
}

\author{\IEEEauthorblockN{Yongjie Wang\IEEEauthorrefmark{1},
Jonathan Cyril Leung\IEEEauthorrefmark{1},
Ming Chen\IEEEauthorrefmark{2}, 
Zhiwei Zeng\IEEEauthorrefmark{3},
Benny Toh Hsiang Tan\IEEEauthorrefmark{1}, \\
Yang Qiu\IEEEauthorrefmark{1}, and 
Zhiqi Shen \IEEEauthorrefmark{2}}
\IEEEauthorblockA{\IEEEauthorrefmark{1}Joint NTU-WeBank Research Centre on Fintech, 
Nanyang Technological University}
\IEEEauthorblockA{\IEEEauthorrefmark{2}College of Computing \& Data Science, Nanyang Technological University}
\IEEEauthorblockA{\IEEEauthorrefmark{3}Joint NTU-UBC Research Centre of Excellence in Active Living for the Elderly, 
Nanyang Technological University \\
Email: \{yongjie.wang, jonathan.leung, chen.ming, zhiwei.zeng, bennytanth, qiuyang, ZQshen\}@ntu.edu.sg}
}
\maketitle

\begin{abstract}
The population of older adults is steadily increasing, with a strong preference for aging-in-place rather than moving to care facilities. Consequently, supporting this growing demographic has become a significant global challenge. However, facilitating successful aging-in-place is challenging, requiring consideration of multiple factors such as data privacy, health status monitoring, and living environments to improve health outcomes. In this paper, we propose an unobtrusive sensor system designed for installation in older adults' homes. Using data from the sensors, our system constructs a digital twin, a virtual representation of events and activities that occurred in the home. The system uses neural network models and decision rules to capture residents' activities and living environments. This digital twin enables continuous health monitoring by providing actionable insights into residents' well-being. Our system is designed to be low-cost and privacy-preserving, with the aim of providing green and safe monitoring for the health of older adults. We have successfully deployed our system in two homes over a time period of two months, and our findings demonstrate the feasibility and effectiveness of digital twin technology in supporting independent living for older adults. This study highlights that our system could revolutionize elder care by enabling personalized interventions, such as lifestyle adjustments, medical treatments, or modifications to the residential environment, to enhance health outcomes. 
\end{abstract}

\begin{IEEEkeywords}
Digital twin, Aging-in-place, Healthcare, Activity recognition.
\end{IEEEkeywords}

\section{Introduction}
Population aging presents a significant challenge to the social and healthcare systems of numerous countries. According to the United Nations \cite{world2015world}, the aging demographic is not only expanding in size but also in proportion. Projections suggest that by 2030, $1$ in $6$ individuals will be aged 60 years or older worldwide. This demographic trend is particularly pronounced in developed countries. For example, approximately 15.2\% of the populace already aged 65 years and over in 2020 in Singapore \cite{malhotra2019aging,demograph2022}. In addition, there is an increasing desire by older adults to remain in their current homes as they become older, rather than moving in with family members or into an assisted living facility. A survey conducted by Singapore's Housing and Development Board (HDB) in 2018 found that 85.9\% of people aged 65 and above and 70.3\% of people between ages 55 and 64 planned to live in their current homes as they aged, whereas in 2014 these percentages were found to be 80.2\% and 60.9\%, respectively~\cite{singapore2021public}. Thus, within the elderly population, there is a noticeable increase in the number of individuals living alone. In Singapore, the number of residents aged 65 and above who live alone has risen from 8.2\% in 2018 to 10.2\% in 2020 \cite{demograph2022}. 

 Aging-in-place refers to the choice of remaining in one's own home as they grow older. This may be preferred because it allows individuals to maintain their autonomy and independence, or because their home has sentimental value and is familiar to them~\cite{stones2016home}. There are, however, challenges associated with aging-in-place. First, while those who choose to age-in-place are typically healthy enough to do so, an elderly individual's overall health may be impaired. However, despite potential impairments, they may still be able to perform their usual activities \cite{world2015world}. Hence, regular health checks and frequent visits to the doctor are still necessary and essential. Secondly, in the case of an emergency, older residents may need a way to quickly contact someone, such as a doctor or a family member. This is particularly important for older residents who live alone. Thirdly, it is said that health is ``a state of complete physical, mental and social well-being'', and not merely the absence of disease or infirmity~\cite{world1948summary}. Therefore, when assessing an individual's overall health, factors such as socioeconomic status, education, and the physical environment should also be considered, which may be difficult to measure in an aging-in-place setting.

To enable successful aging-in-place, advances in signal processing, Internet-of-Things (IoTs), and machine learning can be leveraged to derive early diagnosis and interventions, enhance protection from health emergencies, and promote better health and well-being~\cite{abernethy2022promise,wang2024hybrid,wang2023flexible}. In particular, the accessibility of devices such as fitness bands and wearable patches facilitates the collection and analysis of physical status, experiences, and daily activities, which further contributes to the development and innovation of modern data-driven and AI-powered digital healthcare systems. Based on these techniques, a digital twin can be created, in which a virtual representation of a physical system is developed. 
The digital twin establishes a continuous communication process between the physical system and virtual representation, where the virtual model is regularly updated using data from the physical system, and actions can be taken to improve the physical system based on analyses of the virtual representation \cite{grieves2017digital}.

Using sensor data and machine learning techniques to support aging-in-place presents several benefits. A sensor-based system that continuously monitors an individual's well-being enables the early detection of diseases and helps determine whether a patient needs to visit a healthcare facility in person or if a virtual visit is sufficient. This can reduce costs for the patient and free up resources for healthcare facilities. For example, an analysis in 2020 reported that the need for emergency room care could be reduced by approximately 20\% by using virtual urgent visits~\cite{bestsennyy2021telehealth}. In addition, the wide array of existing sensors can capture key health determinants, providing clinicians and caregivers with information about individuals' daily lives and their common behaviors to support the delivery of preventive and acute care. For individuals with chronic diseases, sensor monitoring can help them manage their disease, as most chronic disease management occurs outside of the hospital. This data can be used to analyze to identify behavioral risk factors that contribute to chronic disease, resulting in real-time and personalized feedback. Overall, digital tools that collect data outside the clinical setting offer meaningful opportunities to identify risks and engage patients, when thoughtfully designed, equitably deployed, and effectively used.

The use of sensors and machine learning in an aging-in-place ecosystem also poses multiple challenges. The primary concern is the privacy and security of data, given that health information is private and potentially trackable. For example, older adults have expressed concerns about sensors that can detect specific motions, record video or audio, or are installed in certain locations such as the bathroom~\cite{jo2021elderly}. However, older adults are more likely to embrace smart-home technology if their privacy concerns are properly addressed~\cite{facchinetti2023can}. Secondly, low-quality data can result from inaccurate devices, environmental noise, and sensor variations, potentially resulting in false alarms that may cause unnecessary anxiety and stress for older people with multiple health conditions~\cite{lomborg2020interpretation}. To address these weaknesses, the appropriate use of digital devices in data collection is crucial. To minimize the damage caused by potential data breaches, we must gather and utilize minimal and necessary information. In addition, it is essential to establish robust monitoring, measurement, and evaluation systems that ensure the effectiveness and reliability of the evidence-based decision-making health monitoring system.

To facilitate successful aging-in-place, in this work, we designed and implemented a sensor system that can be installed in a home to create a \textit{digital twin} of actual events and activities that occurred within the home, as well as conditions of living environment. In particular, to safeguard residents' privacy, our system employs unobtrusive sensors, including motion sensors and low-resolution thermal sensors, which are incapable of collecting identifiable data. This design specifically excludes invasive sensors such as video cameras and microphones. Based on the sensors installed, we implemented machine learning models to make inferences about the resident's well-being, such as in-house activities and living conditions monitoring, which are then shared with healthcare providers and family members through our developed mobile applications. We have deployed our system in two homes and, over a period of two months, we continuously collected data and made inferences and observations using the data to demonstrate the prospective benefits of our system. A large-scale deployment is currently under review and will be tested in the near future.

The rest of the paper is organized as follows: Section \ref{sec:related_work} discusses related work; Section \ref{sec:system_setup} describes our system setup; Section \ref{sec:digital_twin_system} presents our machine learning modules; Section \ref{sec:exp_results} illustrates the use cases of our system; and the final section concludes the paper and outlines future work.

\section{Related Work}
\label{sec:related_work}
Recent advancements in artificial intelligence and digital sensors offer promising use cases for continuous patient monitoring \cite{chen2023framework,lin2019ten}. AI models can be leveraged to predict patients' risks based on sensor data in the early stage, allowing healthcare providers to adjust a patient's treatment based on the analysis from continuous monitoring. For example, smartphone-based photoplethysmography was proposed to detect diabetes using recorded video of a subject's index fingertip \cite{avram2020digital}; digital inhaler sensors have been used to monitor when and where patients with asthma used medications and needed adjustment to treatment plans, reducing the rescue inhaler use and improving the symptom-free days for individuals \cite{merchant2018impact}. 

In aging-in-place, an older adult's performance on Activities of Daily Living (ADLs) is a key indicator of their capability for independent living. ADLs include key routines that should be possible to complete without assistance, such as feeding, dressing, and toileting~\cite{edemekong2019activities}. Krishnan et al. ~\cite{krishnan2014activity} placed passive infrared (PIR) sensors in a home environment to determine a resident's location and used machine learning to predict ADLs. Kaye et al. ~\cite{kaye2011intelligent} presented large-scale and longitudinal studies of unobtrusive sensor systems, which included 265 participants and were conducted over an average period of 33 months. Their systems included PIR sensors to detect location and walking speed, and wireless magnetic contact sensors to detect the opening and closing of doors, thereby inferring time spent outside of the home. Aramendi et al. \cite{aramendi2018automatic} used unobtrusive sensors to collect data from 38 homes of older adults and proposed a machine learning model to first predict ADLs from PIR sensor data. Consequently, their model predicts a person's Instrumental ADL - Compensatory (IADL-C) score and subscores, estimating the difficulty a person has in performing activities such as cooking or cleaning~\cite{schmitter2014development}. Similarly, Alberdi et al.~\cite{alberdi2018smart} proposed a method for predicting symptoms related to Alzheimer's Disease from sensor data. Their model predicts a person's scores on various tests, such as the Timed Up and Go (TUG) and Digit-Cancellation tests, using data from PIR sensors. Their work demonstrates the correlation between sensor data and some test scores, such as TUG speed and overnight movements like toileting~\cite{alberdi2018smart}.

\section{System Setup}
\label{sec:system_setup}
\begin{figure}
\centering
    \includegraphics[width=0.8\linewidth]{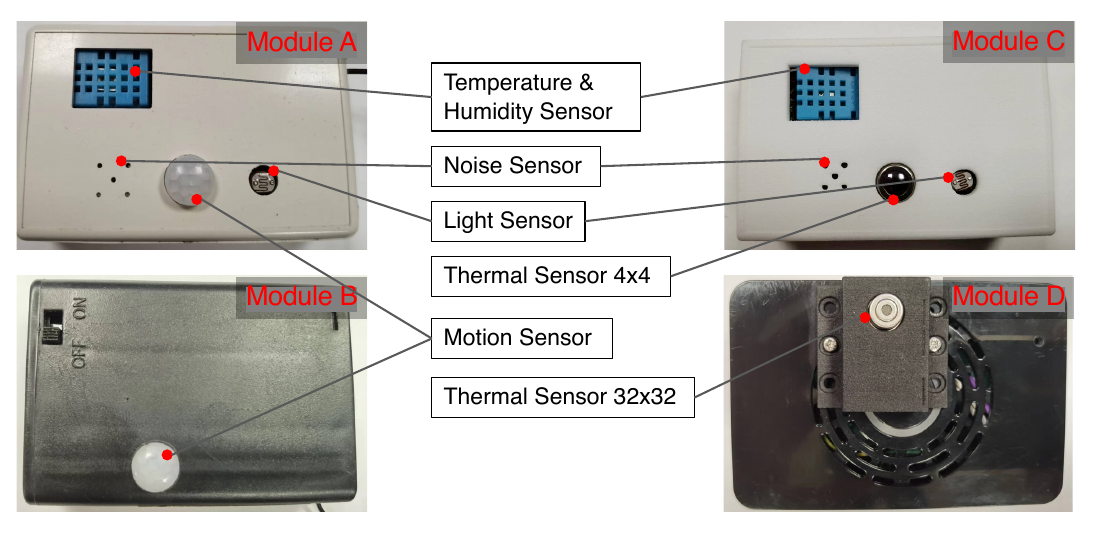}
    \caption{The design of sensor modules.}
    \label{fig:sensors}
\end{figure}

To effectively monitor daily activities of older adults within their homes, we installed certain types of sensor modules throughout the living space, \textit{following the explicit consent of the residents}. The fundamental design principle of this system is to balance (1) the protection of resident privacy with (2) the acquisition of data at a granularity sufficient to support reliable activity monitoring. This principle informs the selection of sensors, as well as the design and configuration of the sensor modules. 

\subsection{Introduction of Sensor Modules}
In terms of sensor selection, we deliberately avoid utilizing potentially invasive sensors such as cameras and microphones. Instead, our system incorporates six types of low-cost sensors: (1) temperature and humidity sensors, (2) noise sensors, (3) light sensors, (4) motion sensors, and (5) $4\times4$ and (6) $32\times32$ resolution thermal array sensors. Temperature/humidity, light, and noise sensors gather data pertinent to their respective environmental parameters. The motion sensor detects physical movements within its range, operating on binary signals. The signal $1$ means that an object is moving within the sensing area of the motion sensor while $0$ indicates that there are no objects moving in the sensing area. Although these sensors provide essential environmental and motion data, their coarse granularity limits their utility for detailed activity inference. To address this, we employ thermal infrared sensors in two resolutions: one with a resolution of $4\times4$ and another with a resolution of $32\times32$. Thermal infrared sensors use the radiation emitted by objects to establish a temperature differential between the object and a heat absorber. This mechanism facilitates the computation and generation of heat maps at varying resolutions. Notably, thermal array sensors are capable of detecting activity even in conditions of low-light or no-light. The data derived from these sensors are simpler to analyze and entail lower overhead complexity compared to image data. Consequently, thermal array sensors were selected for their informativeness, cost-effectiveness, and privacy-preserving attributes. With the six aforementioned sensor types, we designed four sensor modules, labeled from $A$ to $D$ (refer to Figure~\ref{fig:sensors}), each integrating multiple sensors to perform a set of functions. For example, Module $A$ contains four types of sensors: a temperature/humidity sensor, a light sensor, a motion sensor, and a noise sensor. 

\subsection{System Architecture}
\begin{figure}
    \includegraphics[width=\linewidth]{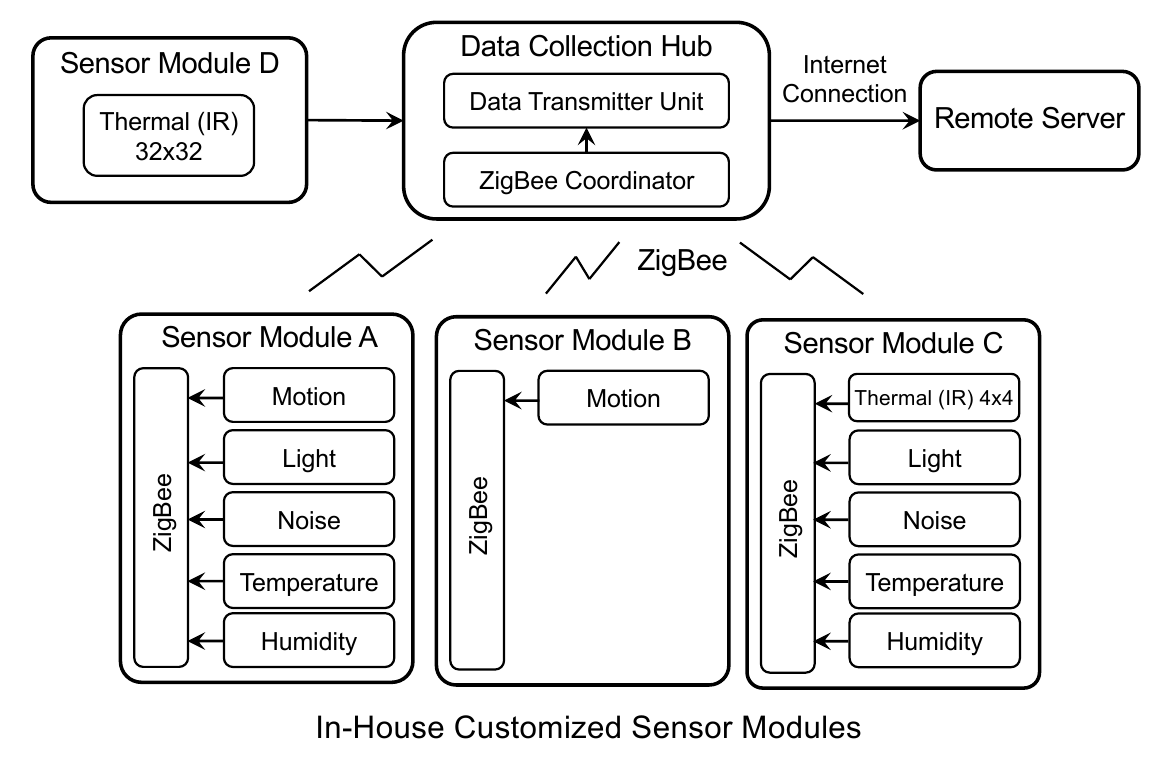}
    \caption{The system architecture for data collection.}
    \label{fig:architecture}
\end{figure}

The system's architecture, supporting data collection across an entire house, is illustrated in Figure~\ref{fig:architecture}. The data collection hub receives sensor signals via the ZigBee protocol and transmits the raw data to a cloud server. Specifically, a ZigBee coordinator within the hub is connected to a Raspberry Pi, which acts as a redirector, serving as an intermediary between the sensors and the cloud. This redirector communicates with the ZigBee coordinator via a USB COM port. It gathers data from the ZigBee coordinator, then packages and transmits data to the cloud server every minute. Each sensor module contains a ZigBee end device that collects sensor signals and wirelessly transmits them to the coordinator. However, Module $D$ diverges from this configuration due to the substantial data volume from the $32\times32$ thermal array sensor, which necessitates a direct connection to the data collection hub via an I2C bus. The cloud server processes the raw data into a feature vector that includes sensor ID, date and time, and value metrics, which are then stored in a MySQL database. The server also directs the thermal array sensors to initiate self-calibration when significant environmental temperature changes are detected and no persons are present within the sensing area. This procedure is designed to mitigate the impact of environmental heat noise and enhance the accuracy of the thermal data.

\subsection{Installation Layout}
\begin{figure}
\centering
    \includegraphics[width=0.8\linewidth]{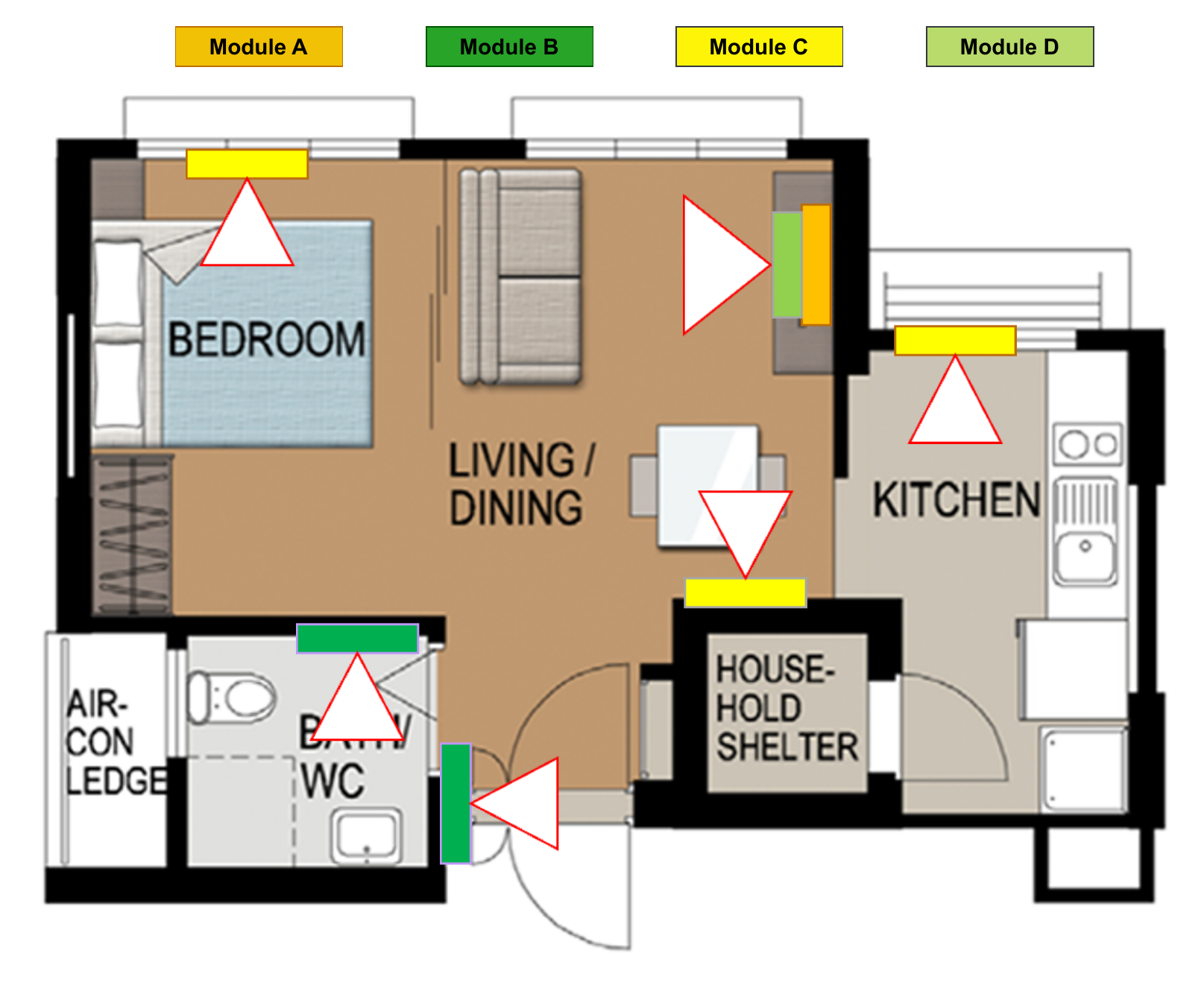}
    \caption{The layout of sensor modules for a 1-bedroom flat. The location of sensor modules can be adjusted based on the flat's layout and room size.}
    \label{fig:layout}
\end{figure}

The installation of sensor modules is carefully planned to ensure comprehensive coverage of areas frequented by the resident, while avoiding the collection of private or identifiable data. 
Figure~\ref{fig:layout} depicts an exemplar layout for a 1-bedroom flat. Two sensor modules $B$ are strategically positioned at the main door and the washroom to monitor entry and exit patterns. Additionally, three Module $C$ units are placed within the bedroom, dining room, and kitchen to track movements and gather environmental data. To uphold data privacy, a low-resolution $4\times4$ thermal sensor (Module $C$) is installed in both the bedroom and kitchen. Conversely, the living room is equipped with a high-resolution $32\times32$ thermal sensor (Module $D$), alongside a Module $A$. It is important to note that the placement of these modules can be tailored to different flat layouts, and room dimensions, or in accordance with resident preferences. 

\subsection{Summary}
Overall, our system has the following advantages: 
\begin{enumerate}
    \item \textbf{Low-carbon footprint}. Our digital solution comprises low-resource devices that consume minimal electricity, demonstrated by the typical household's monthly usage of just $3.96$ kWh. 
    \item \textbf{Full-day monitoring}. Our system provides continuous monitoring throughout the day, ensuring comprehensive coverage and consistent data collection.
    \item \textbf{Privacy preservation}. The system is designed to protect privacy by not collecting personal identifiers, such as facial features or fingerprints, thereby preventing potential breaches of an individual's identity or private activities.
\end{enumerate}

\section{Digital Twin-Based Activity and Health Monitoring}
\label{sec:digital_twin_system}
In this section, we integrate AI models for unobtrusive activity and health monitoring. Our system is designed in a bottom-up manner, starting with sensor-level recognition and then combining these insights to achieve a comprehensive understanding of high-level activities. Specifically, we implement neural networks to recognize postures from thermal data and apply decision rules to infer in-home activities by integrating posture information with signals from other sensors. Finally, we demonstrate how our system supports health status and living condition monitoring.

\subsection{Posture Recognition from Thermal Data}
Thermal array sensors, which offer higher granularity, thereby enable posture recognition that is not achievable with other sensors used. In this phase, we aim to identify distinct body postures using data from thermal array sensors on a per-room basis, leveraging neural networks. Given that there are no public thermal datasets for the devices used in our system, we collected our own datasets for training a neural network model to perform classification of body postures.

\subsubsection{Data Collection} Our dataset was captured in a laboratory setting, utilizing a mock-home environment equipped with our sensor system. Four human participants were involved, and we included 5 different postures in our dataset: `sit', `stand', `walk', `lie down', and `not here'. Each participant was first instructed to maintain each posture for an extended period in a natural manner and then to smoothly transition between different postures. To diversify the data patterns associated with each posture, subjects were guided to vary their distance from the sensors and to position themselves at different angles and places in the sensor range. For walking, participants altered their directions and paces randomly. We also included intervals with no subjects present in the sensing areas. To ensure a uniform sample distribution across all categories, we collect a similar amount of data for each posture.
We stack successive $20$ frames of thermal data ($5$ seconds), as a data sample and annotate it with one of the five aforementioned postures. Finally, the dataset contains $8,646$ high-resolution data samples and $29,840$ low-resolution data samples. We refer to this dataset as the \textit{adult posture dataset}.  

To test our model in real-life setting, we also collected an \textit{elderly posture dataset} as our test data. Six older adults were recruited to perform a pre-defined activity sequence for about 30 minutes in three different houses. Our researchers recorded the corresponding posture labels using a logging tool at the same time. After removing noisy and incomplete data, we obtained a final dataset of 1,385 high-resolution data samples and 4,155 low-resolution data samples.

\subsubsection{Posture recognition} 
\begin{figure}[tb]
    \centering
    \includegraphics[width=\linewidth]{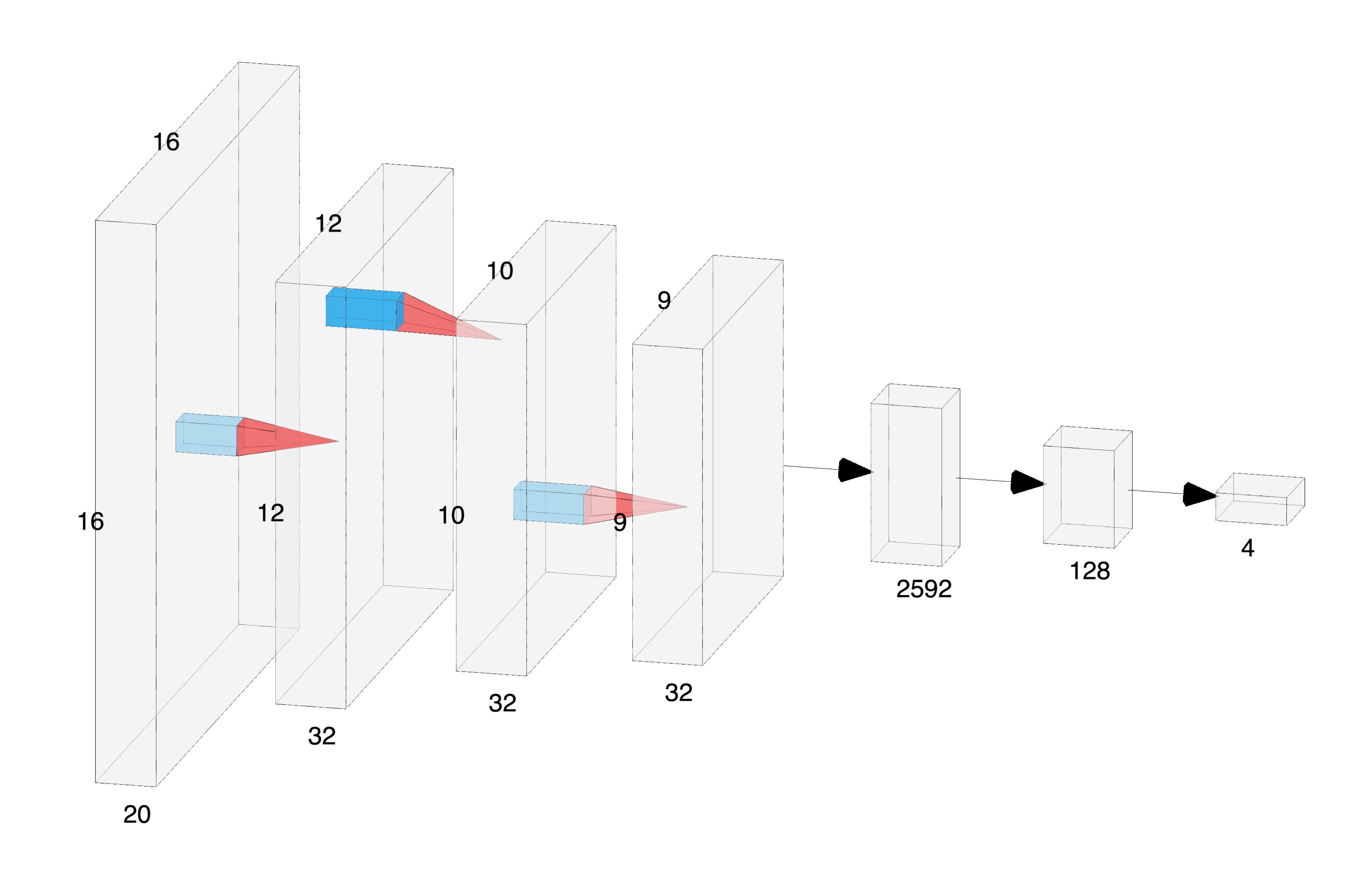}
    \caption{The CNN model architecture used for posture recognition. Batch normalization, ReLU activation, max pooling, and dropout operation are adopted after each convolutional layer. }
    \label{fig:CNN_architecture}
\end{figure}

We first preprocess the thermal data by filtering out persistent environmental heat noise, such as heat patches caused by sunlight. Next, we leverage a customized convolutional neural network (CNN) to capture the spatial-temporal dynamics of human activity. As shown in Figure \ref{fig:CNN_architecture}, our CNN model contains multiple convolutional layers, followed by full-connected layers for classification. We also employ batch normalization and dropout techniques to prevent overfitting. During training, we split the \textit{adult posture dataset} into training and test sets at a ratio of $4:1$. The training data is then further split into training and validation sets at a ratio of $4:1$ too. We trained our CNN model on a GPU cluster on the training set, where the cross entropy loss is minimized with the Adam optimizer. The training process stops until it reaches a predefined number of iterations. We select the model that achieve the best performance on the evaluation set. 

The test accuracy on the test set of the \textit{adult posture dataset} is $95.12\%$ for the high-resolution data and $99.32\%$ for the low-resolution data. In addition, the test accuracy on the \textit{elderly posture dataset} is $84.10\%$ for the high-resolution data and $88.20\%$ for the low-resolution data.  

\subsection{In-house Activity Recognition}
In this phase, we aim to identify seven high-level activities: sleeping, leaving/returning home, kitchen activity, dining room activity, living room activity, using the restroom, and hosting visitors, assessed on a per-household basis. This analysis integrates results from the initial posture recognition in each room, with data from other sensors, historical activity records, and contextual factors such as time. Activity recognition is conducted for data collected in 5-second intervals for each house, considering that meaningful activities generally occur over minute-long periods rather than seconds. 

\subsubsection{Data Collection} This dataset was collected from three different houses with three adult human subjects. After installing the sensor system, each participant executed specific activity sequences, covering all activity classes to be recognized. They were instructed to perform these activities in a manner reflecting their typical behavior for $30$ minutes. Concurrently, our researchers recorded both the posture labels and activity labels with a logging tool. Each participant repeated thrice for each activities, yielding about $360$ minutes of data per house. The data is associated with one of seven classes: `sleeping', `kitchen activity', `dining room activity', `not at home', `restroom', `living room activity', and `visitors'. We refer to this dataset as the \textit{adult ADL dataset}.

During the collection of the \textit{elderly posture dataset}, three elderly participants were instructed to perform specific activity sequences in a manner akin to the methodology employed for the \textit{adult ADL dataset} collection. This yielded an \textit{elderly ADL dataset} containing $115.4$ minutes of data. 

\subsubsection{Activity recognition} The activity classification is primarily based on a priority-rule-based algorithm, which integrates posture recognition results from each room over one minute, along with inputs from other sensors, historical activities, and contextual information. When posture recognition in multiple rooms indicates the inhabitant's presence, the key principle is to consider the room with the highest priority for activity recognition. Restroom activity receives higher priority because its binary motion sensor trigger is less susceptible to noise, and restroom usage patterns are important for health monitoring. In rooms equipped with thermal array sensors \textemdash such as the living room, dining room, kitchen, and bedroom \textemdash their priorities are determined by a derived motion index. This index quantifies changes in heat patterns to distinguish between human activity and environmental heat noise. The priority is also contingent on the context. For example, the bedroom will be given higher priority during the nighttime. The recognition of some activities is also dependent on activity history. For example, the `not at home' activity is recognized in a post-hoc manner over a longer time frame, identifying patterns with two paired doorway motion sensor triggers and no intervening activities in any of the rooms. The accuracy of our priority-rule-based algorithm on the \textit{adult ADL activity} dataset is 73.9\%, and on the \textit{elderly ADL dataset} is 78.2\%.

\subsection{Health and Environment Monitoring}
From activities occurring within the household, our system can detect several unusual patterns that may be indicative of specific health conditions or suggest living environments detrimental to well-being. The key insights are as follows:

\textbf{Sleep duration.} Utilizing our activity recognition model, we detect sleep activity in the bedroom and calculate its duration to estimate sleep hours. Since the thermal sensor is sensitive to the residual heat from the area where the person was lying, we leverage
the aforementioned priority-based algorithm and motion index to reduce false predictions of sleeping.  In particular, sudden changes in lighting detected by the bedroom's light sensor assist in discerning non-sleep-related activities. For instance, by integrating posture recognition results with sudden light changes and motion signals from the restroom, we can estimate urination frequency during the night.
Both long-term sleep deprivation and prolonged sleep should raise the awareness of caregivers or designated family members.

\textbf{Sleep quality.} From the thermal sensor data in the bedroom, we can analyze the frequency and intensity of body movements on the bed to infer sleep disturbances or wakefulness periods, which serve as coarse indicators of sleep quality. Specifically, we measure the differences between two consecutive data frames. If the difference is below a certain threshold, we recognize that the elderly individual remains still; otherwise, movement is detected. This method allows for an estimation of sleep quality for elderly participants. It is critical to note that calibration is necessary to establish an appropriate threshold for accurate assessments. 

\textbf{Living environment.} Noise pollution from construction and traffic is a pervasive issue in Singapore due to its dense urban environment, potentially affecting sleep quality and other daily activities of older people. Our system includes noise sensors that could record the environment noise, temperature and humidity sensors that regularly record the bedroom's temperature and humidity. Given Singapore's tropical climate, characterized by high humidity and temperatures throughout the year, there is an elevated risk of respiratory and cardiovascular diseases. It is imperative to be vigilant and take measures to improve living conditions if residents are exposed to such adverse conditions for extended periods.

\textbf{Time spent outdoors.} Loneliness among the elderly is a pressing concern with profound implications for individual well-being and societal health, especially for those living alone. Loneliness significantly impacts mental health, increasing the risk of depression and anxiety, as well as physical health problems like cardiovascular disease and a weakened immune system.
Our system can identify the `not at home' status by detecting no activity across all rooms while recording a trigger at the door's motion sensor. Similarly to monitoring sleep duration, we track the time spent outside the home, and if necessary, the system can alert caregivers or family members to check in on the resident.

\begin{figure}
    \centering
    \includegraphics[width=0.7\linewidth]{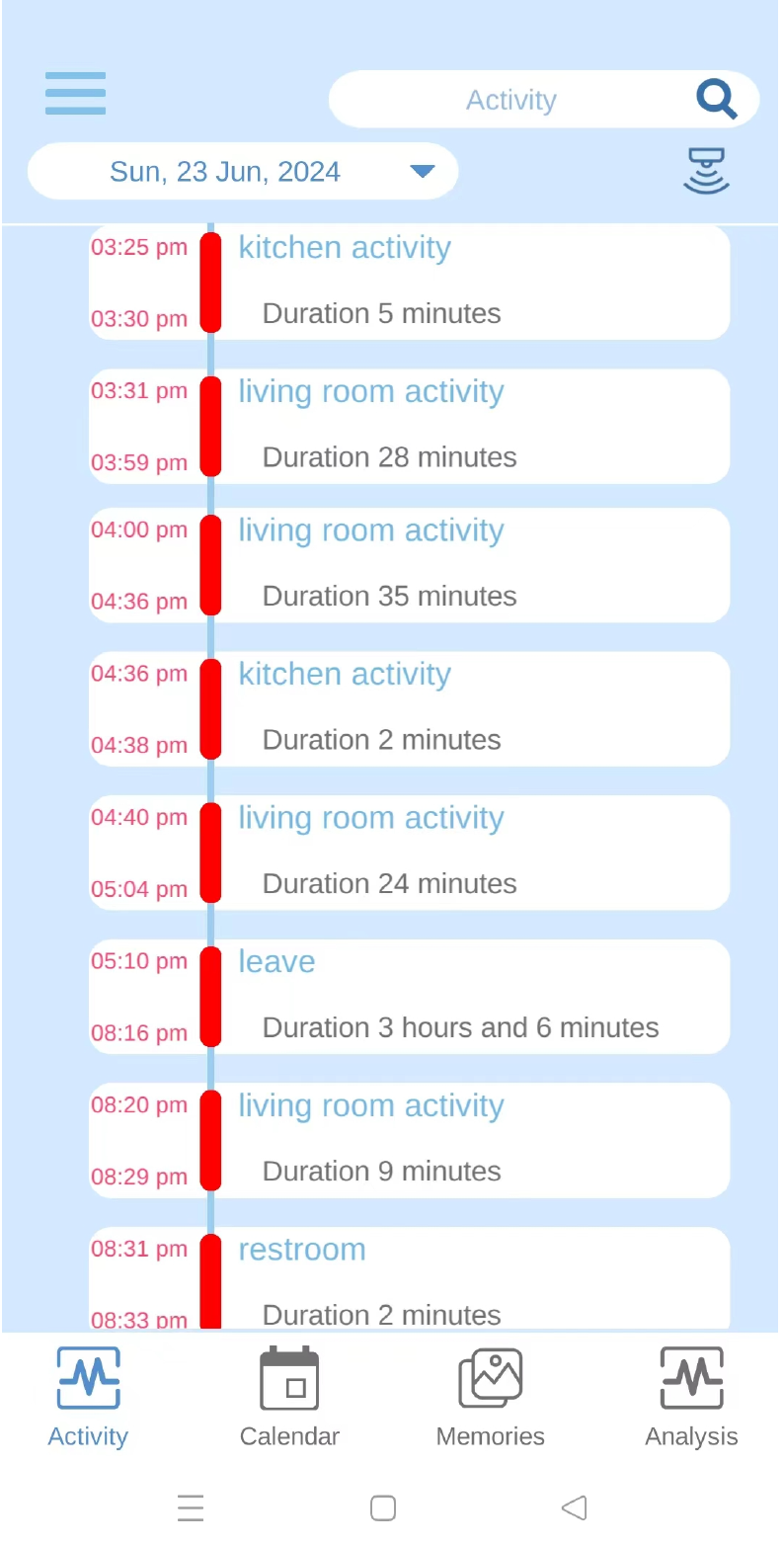}
    \caption{The screenshot of our developed mobile application.}
    \label{fig:mobile}
\end{figure}

We have developed a mobile application, as shown in Figure \ref{fig:mobile}, which integrates the aforementioned functions into a comprehensive platform for elderly health monitoring. This application displays the daily activities and living conditions of the elderly, allowing designated caregivers and family members to monitor their profiles. Based on our data analysis, the application offers real-time insights and alerts, ensuring timely interventions when necessary. 


\section{Experimental Results}
\label{sec:exp_results}
We deployed our system in two houses to monitor activity within designated areas over a period of two months, after receiving approval from the Institutional Review Board. In total, we collected approximately 60 GB of raw data, which serves as a valuable resource for health monitoring. To validate the feasibility of our system, we report two case studies to demonstrate how our system monitors household activities and living environments.

\subsection{Sleep Duration Analysis}

\begin{figure*}
    \centering
    \includegraphics[width=0.95\linewidth]{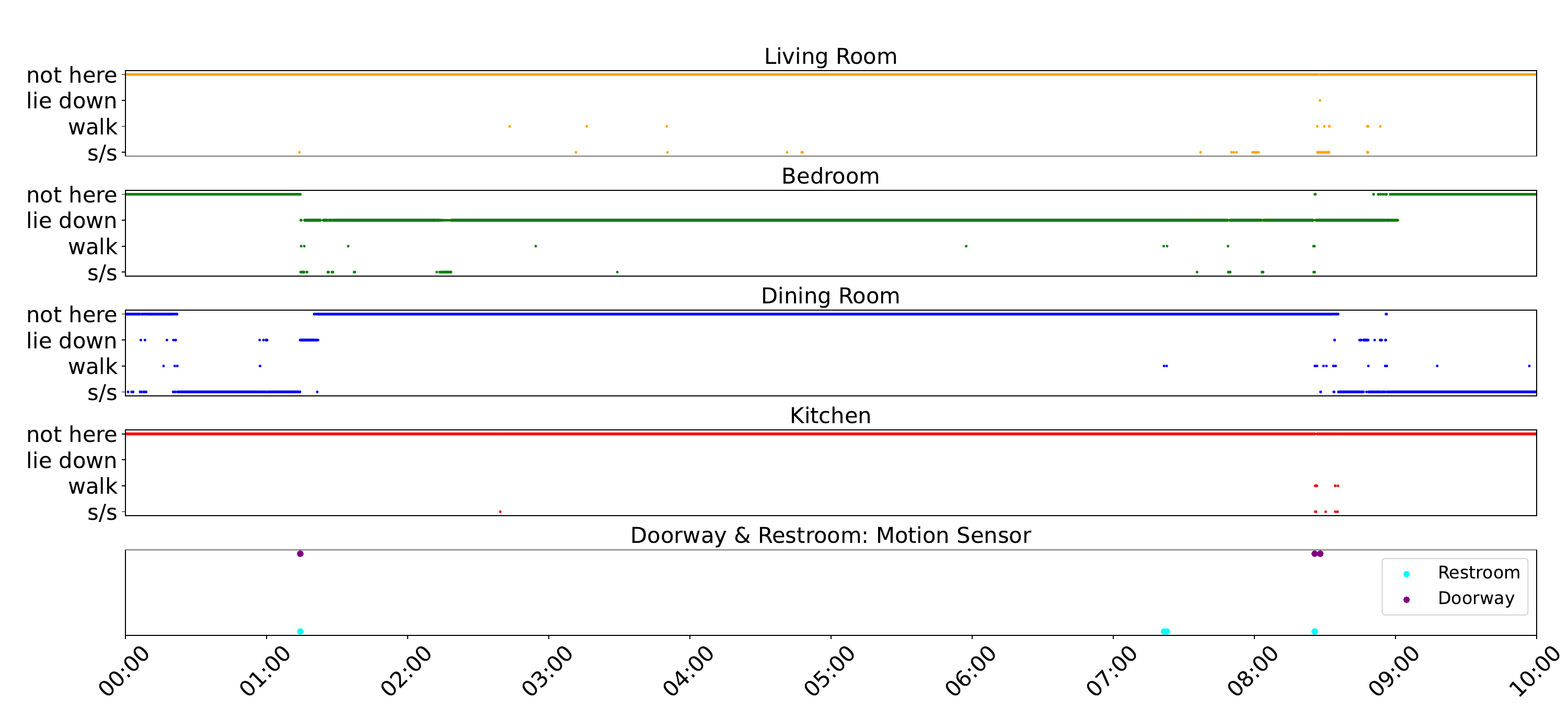}
    \caption{The sleep duration analysis. Activities occurring outside the bedroom are used to accurately measure the sleep duration. }
    \label{fig:sleep_monitoring}
\end{figure*}

Our system performs sleep duration analysis by inferring postures and activities occurring in different rooms using our machine learning models. The analysis indicates that the participant remained seated in the dining room until about 1:00 AM, visited the toilet, and then went to bed, sleeping for approximately six and a half hours. After a subsequent visit to the toilet, the participant laid down and slept for an additional hour. Notably, from 8:30 AM to 9:00 AM, residual heat on the bed led to a misclassification to `lie down' posture in the bedroom. To address such errors, we consider the motion indices in the various rooms and other context rules (e.g. current time) to determine the priority of activity recognition. In the current scenario, a high motion index in the restroom and dining room is used to signal potential departure from the bedroom, under the assumption that the participant lives alone. This data allows us to aggregate daily sleep durations to identify potential sleep deprivation or excessive sleep issues.

In a similar manner to what is plotted in Figure \ref{fig:sleep_monitoring}, we can evaluate the time the resident adult spends outside the household by detecting no activity across all rooms in conjunction with the triggering of the door's motion sensor. This duration spent outside can be utilized to assess the potential loneliness of the elderly resident.

\subsection{Living Environment Report}

\begin{figure*}
    \centering
    \includegraphics[width=0.95\linewidth]{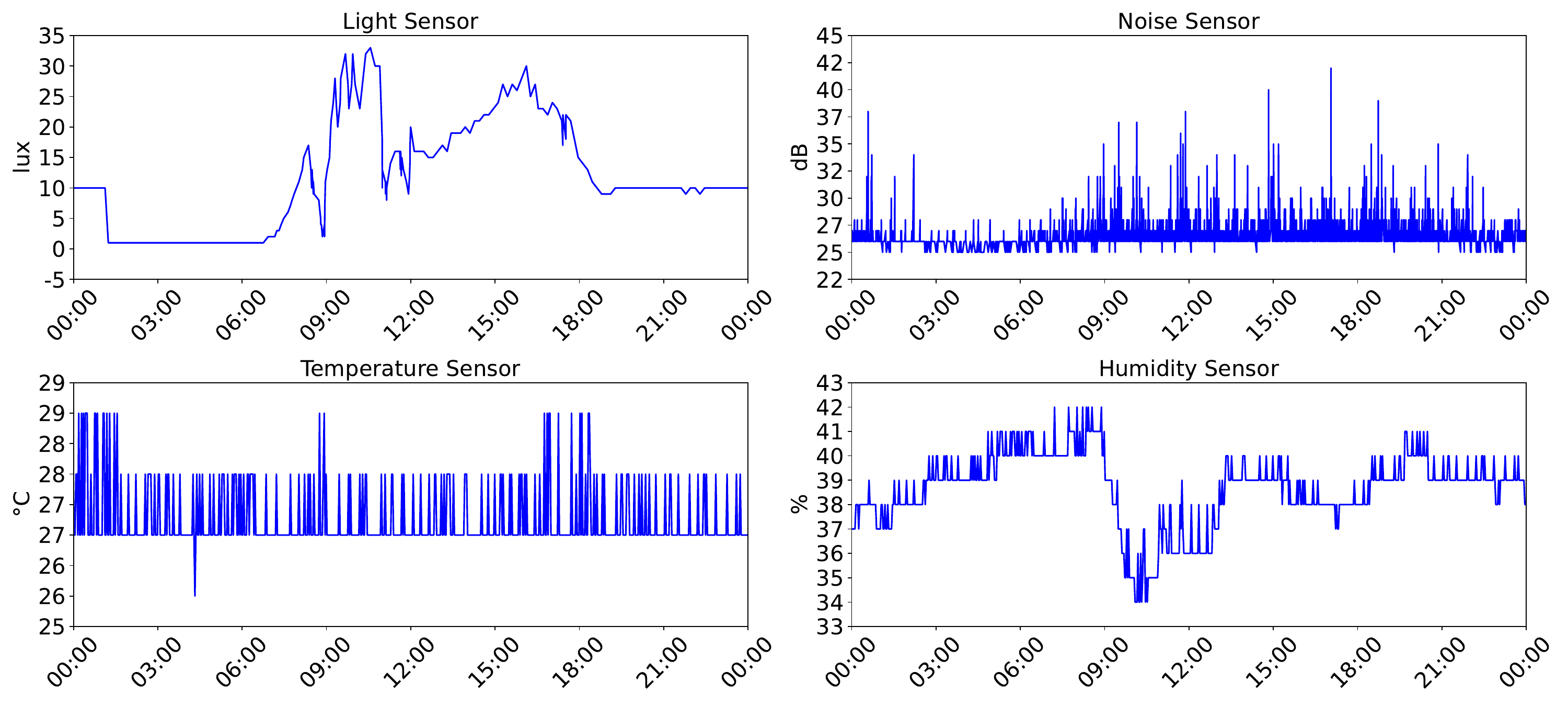}
    \caption{The environment monitoring on dining room.}
    \label{fig:environment}
\end{figure*}

In this study, we report the environmental data collected by our sensors over a 24-hour period, as shown in Figure \ref{fig:environment}. The figure includes readings from the light, noise, temperature, and humidity sensors. Light sensor readings reveal the typical cycle of natural daylight and changes indicative of artificial lighting activation. Sudden changes in light or temperature are also instrumental in facilitating activity recognition. The noise sensor records the noise levels, with higher intensity during the afternoon and early evening. Overall, the data from these sensors provides a comprehensive view of the environmental conditions within the monitored area, which is crucial for assessing the living environment and its impact on the elderly. By analyzing these environmental patterns, we can make informed adjustments to enhance the well-being and comfort of the residents. 

\section{Conclusion and Future Work}
\label{sec:conclusion}
Population aging presents a significant challenge to the public health system. In this work, we propose an unobtrusive sensor system that can be installed in the homes of older adults to facilitate aging-in-place while preserving the privacy of residents. By installing various sensors within their living areas, we build a digital twin of residents' activities and home environment. We leverage machine learning models to analyze their daily activities and living environments, empowering older adults to live independently as they desire. We have deployed our system in two homes and continuously monitored the residents' activities over a period of two months. Our system demonstrates a successful use case of digital twin technology in assisting with aging-in-place. In future, we will install our system in more homes of older adults and collect more data to calibrate and improve our system. 

However, there are several challenges that remain to be addressed in future work. First, our system assumes a single resident at home, but it should be capable of interpreting and managing scenarios involving multiple occupants or visitors. Second, the system relies on neural network models, whose decision-making process difficult to interpret due to their inherent opacity. Extensive efforts are needed to understand their internal mechanisms to enhance the trustworthiness of the system \cite{wang2024gradient}. Lastly, large-scale user experiments remain challenging due to the need for technicians to install the system in various home layouts, as well as the time-consuming process of obtaining both governmental approval and household consent. Additionally, the data collection process should include participants from diverse ethnic and cultural backgrounds to mitigate potential biases.

\bibliographystyle{IEEEtran}
\bibliography{ref}
\end{document}